# Interfacial Entropic Interactions Tunes Fragility and Dynamic Heterogeneityof Glassy Athermal Polymer Nanocomposite films

Nafisa Begam[1], Nimmi Das[1], Sivasurender Chandran[2], Mohd Ibrahim[1], Venkat Padmanabhan[3], Michael Sprung[4], J. K. Basu[1]

[1] Department of Physics,Indian Institute of Science Bangalore, 560012,India

[2] Institute of Physics, University of Freiburg, 79104 Freiburg, Germany

[3] Department of Chemical Engineering, Tennessee Technological University,Cookeville,TN 38505,USA

and

[4] Deutsches Elektronen Synchrotron DESY, Notkestresse 85, 22607 Hamburg, Germany

Enthalpic interactions at the interface between nanoparticles and matrix polymers is known to influence various properties of the resultant polymer nanocomposites (PNC). For athermal PNCs, consisting of grafted nanoparticles embedded in chemically identical polymers, the role and extent of the interface layer (IL) interactions in determining the properties of the nanocomposites is not very clear. Here, we demonstrate the influence of the interfacial layer dynamics on the fragility and dynamical heterogeneity (DH) of athermal and glassy PNCs. The IL properties are altered by changing the grafted to matrix polymer size ratio, $f$, which in turn changes the extent of matrix chain penetration into the grafted layer, $\lambda$. The fragility of PNCs is found to increase monotonically with increasing entropic compatibility, characterised by increasing $\lambda$. Contrary to observations in most polymers and glass formers, we observe an anti-correlation between the dependence on IL dynamics of fragility and DH, quantified by the experimentally estimated Kohlrausch-Watts-Williams parameter and the non-Gaussian parameter obtained from simulations.

## 1 Introduction

The nature of nanoparticle (NP)-polymer interface plays a crucial role in determining flow, thermal, mechanical, optical and electrical properties [1-3] of polymer nanocomposites (PNC). Almost a decade back we had shown, for the first time, how glass transition temperature, $T_g$, can be tuned by controlling the nature of NP-polymer interfacial layer (IL) and confinement of polymer chains due to NPs in athermal PNCs [4]. Recently, there has been renewed interest in correlating several other properties of PNCs in terms of the structure and dynamics of IL especially for enthalpic PNCs [5-8]. A fundamental difference between the IL in enthalpic and entropic PNCs, often based on polymer grafted nanoparticles (PGNP), used by us and others [9-12] is in their nature and formation. While the width and structure of the IL for enthalpic PNCs is largely determined by the strength of the NP-polymer segmental interactions and the molecular weight of the matrix polymer [5,6], for PGNP based entropic PNCs this is dependent on the ratio of the grafted and matrix polymer molecular weight, $f$, grafting density and NP size among other parameters [9,10]. More importantly, $f$, also determines the effective entropic interaction between the PGNPs turning from attractive at low $f$ to repulsive at larger $f$ which also allows control over the state of dispersion of PNCs which ultimately determines the efficacy of these materials. At a more fundamental level the matrix chain penetration depth, $\lambda$, is directly proportional to $f$ and can be defined as the microscopic parameter which controls the extent of entropic interaction [13]. The concept of fragility, $m$, has proven to be useful in understanding the diversity of dynamics seen in various glass formers, including polymers [14,15]. Some understanding has emerged to explain the differences in the fragilities and its relation with various thermodynamic properties, physical aging, and mechanical behavior [16,17], while for soft colloidal glass formers it has been shown that fragility can be tuned by varying the strength of the inter-particle interactions [18].

In enthalpy dominated PNCs the fragility was shown to increase with the inclusion of particles with

Table 1: Sample details

| Sample | $W_m$ kDa | $W_g$ kDa | $f$ | $R_c$ nm | $R_g$ nm |
|---|---|---|---|---|---|
| PS50k | 50 | - | 0.0 |  | 5.9 |
| 3k50k | 50 | 3 | 0.06 | 6 |  |
| 20k50k | 50 | 20 | 0.4 | 8.5 |  |
| PS100k | 100 | - | 0.0 |  | 8.45 |
| 3k100k | 100 | 3 | 0.03 | 6 |  |
| 20k100k | 100 | 20 | 0.2 | 8.5 |  |

$W_m$ and $W_g$ are the molecular weight of matrix and grafted chains, respectively; $R_c$ is the size of nanoparticle determined from small angle X-ray scattering (Fig. S1 [27]); $R_g$ is the radius of gyration of matrix chains. Grafting densities of 3k and 20k grafted nano particles are 1.7 and 1.3 chains/nm$^2$, respectively.

attractive interactions with the matrix polymers, while the particles with repulsive interactions show a decrease in the fragility [19]. It has been shown recently, that apart from $T_g$ and viscosity, fragility in enthalpic PNCs can be tuned by changing the structure of the IL [5,6,8,19,20]. However, for entropic PNCs it was suggested that fragility remains largely unaltered due to inclusion of NPs [21].

Another key dynamical parameter, which has been widely used to describe glasses and supercooled liquids, including polymers and PNCs, is dynamic heterogeneity (DH) [22-24]. In the case of enthalpic PNCs, $m$ seems to correlate with DH (increases for attractive and decreases for repulsive interactions) quantified in terms of a temperature dependent dynamical length scale, $\xi$, which seems to grow moderately on approaching the glass transition [5,19,24,25]. On the other hand some studies suggest that DH decreases with increasing fragility [26] in amorphous polymers. Surprisingly, very little is known in terms of fragility, DH or their correlations for entropic PNCs [4,21,23].

In this report, we study the temperature dependent relaxation dynamics of typical entropic PNCs consisting of polystyrene grafted gold nanoparticles dispersed in polystyrene (PS) films. Using the technique of X-ray photon correlation spectroscopy (XPCS), we extract temperature dependent viscosity, $\eta$, of the various PNC films as a function of $f$ and estimate their respective fragilities, $m$. We find that in these PNCs $m$ increases with increasing $f$, which is also corroborated, qualitatively, by coarse grained molecular dynamics (MD) simulations on similar systems. The experimentally extracted Kohlrausch-Watts-Williams (KWW) parameter, $\beta$, is a typical parameter used to quantify DH. $\beta$ (DH) was found to increase (decrease) with increasing $f$. Coupled with this the non-Gaussian parameter (NGP) and interface chain diffusivity was found to decrease with increasing $f$ suggesting a correlation between DH and IL dynamics. However the variation of $m$ and $T_g$ with $f$ showed anti-correlation compared with the variation of DH with $f$, contrary to several earlier studies on glasses and PNCs.

## 2 Experimental details

The results presented in this letter are based on thin films (thickness, $h$, $\sim$ 65-70 nm; Table S2 [27]) of PS of two different molecular weights (100kDa and 50kDa), embedded with PGNPs having two different grafted chain molecular weights (3kDa and 20kDa) synthesized using the methods discussed earlier [28-30]. These PGNPs were dispersed in linear PS at 0.5% volume fraction of the gold core. Details of samples are provided in the supplemental information (SI) [27] as well in our earlier studies [28,29] and summarized in Table 1.

X-ray photon correlation spectroscopy measurements were performed on the annealed PNC films in grazing incidence geometry at PETRA III (beamline P10), DESY, Hamburg, Germany. The measurements were done with an X-ray beam of energy 8 keV and beam size 25×25 $\mu m^2$. The beam was incident on the sample at an angle $\theta$, chosen to be smaller than the critical angle of the films for that particular X-ray energy. The critical angle was estimated as the position of first minima of the X-ray reflectivity profile which appears before the substrate critical angle [31]. Typically, the PNC films critical angle was found to be around $\sim 0.16°$. The scattered intensity from the samples were collected by a CCD detector (Lambda, area 1536×512 $px^2$ (84.5×28.2 $mm^2$), pixel size of 55×55 $\mu m^2$) as a reciprocal image. A typical CCD image of such scattering is shown in Fig. S4 [27]. The measurements were performed at the temperatures varying from 413K to 463K. To obtain the intensity autocorrelation function, $g_2(q_x, t)$, typically, 2000 images were collected with an interval of 0.4 sec at each temperature for all the samples. From these image series, we have extracted $g_2(q_x, t)$ and modeled them by a function of the following form [32-34]

$$g_2(q_x, t) = 1 + b|F(q_x, t)|^2, \qquad (1)$$

where, the intermediate scattering function (ISF) is given by

$$F(q_x, t) = exp-(\Gamma \cdot t)^\beta, \qquad (2)$$

where $b$ is an instrumental factor called the speckle contrast, $t$ is delay time, $\Gamma = 1/\tau$ is the relaxation rate, $\tau$ is the relaxation time and $\beta$, the Kohlrausch exponent, is a canonical gauge for quantifying the deviation in the relaxation behavior from simple exponential.

Wave vector dependence of relaxation time is modeled using hydrodynamic continuum theory [34-36] and relaxation rate is given by [37]

$$\Gamma(q_x) = \frac{1}{\tau(q_x)} = \frac{\gamma}{2\eta} \frac{q_x(sinh(q_xh)cosh(q_xh) - (q_xh))}{cosh^2(q_xh) + (q_xh)^2}. \qquad (3)$$

Here, $\eta$ and $\gamma$ denotes the viscosity and surface tension of the film, h and $q_x$ are thickness of the film and in-plane wave vector respectively.

The viscosity of the film shows temperature dependence which is modeled using well known Vogel-Fulcher-Tammann (VFT) equation [38],

$$\eta = \eta_o \exp\left(\frac{BT_o}{T - T_o}\right), \qquad (4)$$

where $\eta_o$ and $B$ are the fit parameters, $T$ is absolute temperature and $T_0$ is VFT temperature [27].

Using the temperature dependence of viscosity as obtained from eqn. 4, the fragility, $m$, can be estimated [27,39] using the definition,

$$m = \frac{\partial ln\eta}{\partial \left(\frac{T}{T_g}\right)}\bigg|_{T_g} \qquad (5)$$

where $T_g$ is the glass transition temperature of the films.

Figure 1(a) shows representative intermediate scattering functions (ISFs), $F(q_x, t)$, collected from XPCS measurements for PS50k based PNCs at 433 K (Fig. S6 [27]). In comparison with neat polymer, a faster dynamics can be observed for 3k50k ($f = 0.06$) sample, while 20k50k ($f = 0.4$) sample shows a slower dynamics. Similar behavior can be observed for PS100k based samples as shown in Fig. 1(b). To quantify the observed variations, we have extracted relaxation rate from the ISF (described in SI [27]) by modeling the ISF with eqn. 1 in SI with relaxation time, $\tau$, Kohlrausch exponent, $\beta$ and instrumental contrast $b$ as fit parameters. The relaxation rate ($\Gamma$) as obtained for PS50k based samples is summarized in Fig. 2. Data for PS100k based samples is shown in Fig. S7. Figure 2(a) shows the wave vector dependent relaxation rate for neat PS films (PS50k) for different temperatures. As expected for a normal viscous melt, the dynamics becomes faster with increasing temperature [31,34]. The wave vector dependence is suggestive of capillary wave dynamics and we have used the well established formalism [34-36] to extract viscosity from such data using the eqn. 3. Using the well known temperature dependent $\gamma$ values for PS [40,41], we have extracted the viscosity, $\eta$, for all samples at the corresponding temperatures, using eqn. 3.

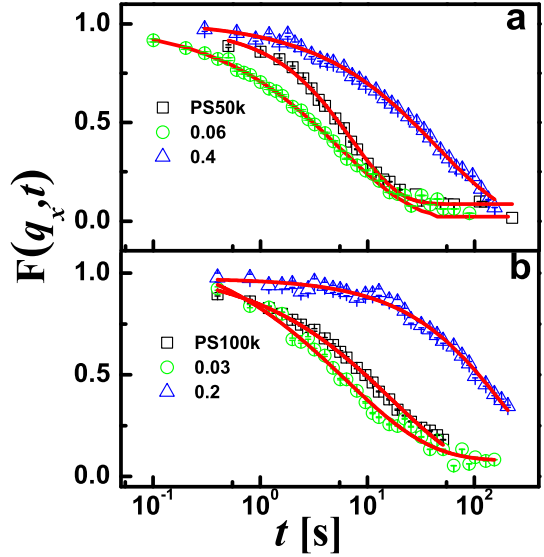

Figure 1: Representative ISF (a) for pure PS50k and corresponding composites with $f = 0.06$ and $f = 0.4$ samples and (b) for pure PS100k, $f = 0.03$ and $f = 0.2$ as indicated in respective panels. All the correlation functions were obtained at temperature 433 K and the lateral wave-vector $q_x = 7 \times 10^{-4} Å^{-1}$. The solid red lines are the fit to the eqn 2.

Extracted values of viscosity for PS films agrees well with the literature values of corresponding bulk PS viscosity as a function of temperature as can be seen in Fig. 3 (a) [42]. The validation of this technique of estimating film viscosity has been well established by several groups [34,43]. The viscosity obtained from XPCS has been found to be in good agreement with the bulk PS viscosity up to a very high molecular weight [36]. In addition, XPCS provides information about dynamic heterogeneity, length scale dependent dynamics and structural variations underlying the dynamics in an integrated manner. Therefore, XPCS is a technique of choice at this stage to acquire all such thin film information.

Figure 3(b-c) summarize the temperature dependent viscosity behavior for all the samples. For both PS50k (Fig. 3(b)) and PS100k (Fig. 3(c)) based mixtures, it can be observed that the $\eta$ of PNCs with smaller $f$ is comparably lower than the neat polymers, while the samples with higher $f$ show a slight increase in $\eta$. Such reduction in the viscosity of PNC was observed earlier [7,44-46] and alluded to the presence of thin interface layer surrounding the nanoparticles with reduced surface viscosity [7,44] Our earlier observations [31] also indicated the presence of an IL with reduced viscosity compared to the bulk. To quantify the temperature dependence of the observed viscosity, we have modeled the viscosity

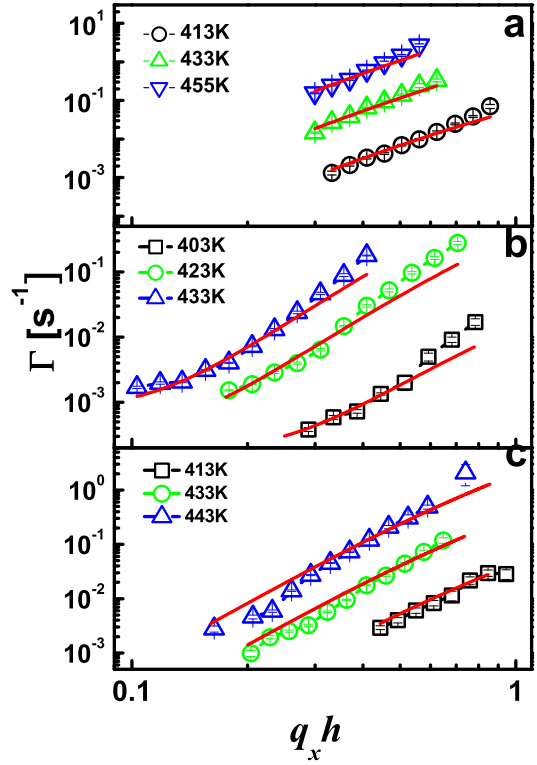

Figure 2: Wave vector ($q_x$) dependent relaxation rate $\Gamma$ as a function of temperature for PS50k (a), 3k50k (b) and 20k50k (c). The red dashed lines are the fits to the eqn 3.

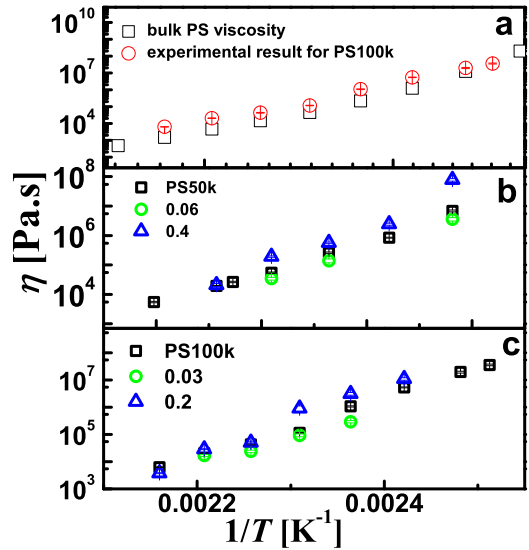

Figure 3: (a) The comparison of the viscosity estimated for PS100k films and literature values of PS94k bulk viscosity [42]. The Comparison of extracted viscosity, $\eta$, from capillary wave fit (eqn. 3) for PS50k matrix based (b) and PS100k matrix based PNCs (c) as a function of $T$ for various values of $f$. The reduction in $\eta$ is evident for smaller $f$ and comparatively larger viscosity for higher $f$.

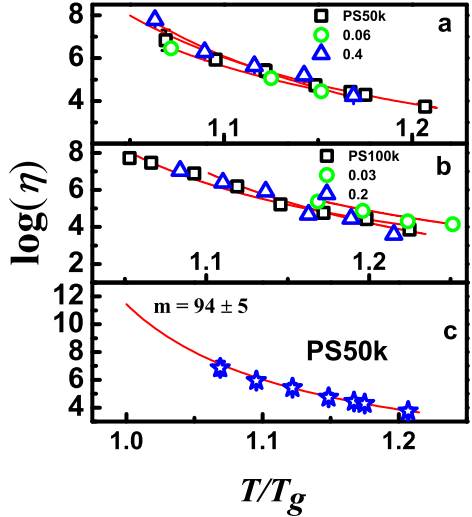

Figure 4: The viscosity as a function of $T/T_g$ for PS50k (a) and PS100k (b) based samples are shown along with the VFT fits (solid red lines) and, (c) a typical extrapolation of the VFT fit from which the slope near $T_g$ ($T/T_g = 1$) is calculated to determine the fragility, $m$.

plots with VFT equation as given in eqn. 4 . The fragility, $m$, is calculated using eqn. 5. $T_g$ of the films used for the fragility estimation is estimated from the technique AFM force-distance spectroscopy. Since normal calorimetric methods like differential scanning calorimetry (DSC) can not be used for this films, we have employed this technique which has been well established earlier by ourselves and others [28,47,48] and found reliable to estimate $T_g$ for this type of ultra-thin films. A comparison of $T_g$s estimated from AFM and that from differential scanning calorimetry (DSC) on several films are shown in table S13 in SI [27]. This suggest that the $T_g$ obtained from AFM is quite reliable.

## 3 Results and discussion

Figure 4(a-b) shows the viscosity of all PS50k and PS100k based samples as a function of $T/T_g$ with corresponding VFT fits. In order to obtain the fragility, we have extrapolated this fit till $T/T_g = 1$ and taken the slope at that point which is demonstrated in Fig. 4(c) for pure PS50k data. Figure 5(a) summarizes the observed variation of $m$, normalized with that of corresponding pristine polymers ($m_{PS}$), with $f$. The fragility shows an overall trend of lower values at lower $f$ which rises to values comparable to or even higher than the pristine PS at larger $f$. In order to check the validity of this observation, we re-calculated the errors in $m$ using the methods described in SI [?] which includes errors in all the parameters and provides a reliable error bars.

Most reports suggest a direct correlation between the changes in $T_g$ and $m$ [20,49]. Our experimental results for change of $T_g$ (Fig SI 13) and $m$ with $f$ also follows similar trends. However, it must be highlighted that the extent of fragility change is much more significant than that of corresponding change in $T_g$. This signifies that there could be an additional effect of the entropic interaction parameter $f$ on the fragility.

In order to obtain microscopic insight on the effect of $f$, on $m$, we have estimated $\eta$ of a system with linear polymer chains embedded with grafted nanoparticles which is similar to the experimental system using MD simulation following established methods [50]. Here grafted chain length was varied keeping matrix same to get a range of $f$ values [27]. The temperature dependent viscosity obtained from these simulations (Fig. S11) has been used for calculating fragility as shown in Fig. 5(b). Interestingly, $m/m_0$ shows a significant reduction at small $f$ as observed experimentally for our systems (Fig. 5(a)). At larger $f$, $m$ increases similar to our experimental observations although it does not exceed the value of pure PS. Thus a fairly good qualitative correlation between the experimental and simulation results for fragility is observed.

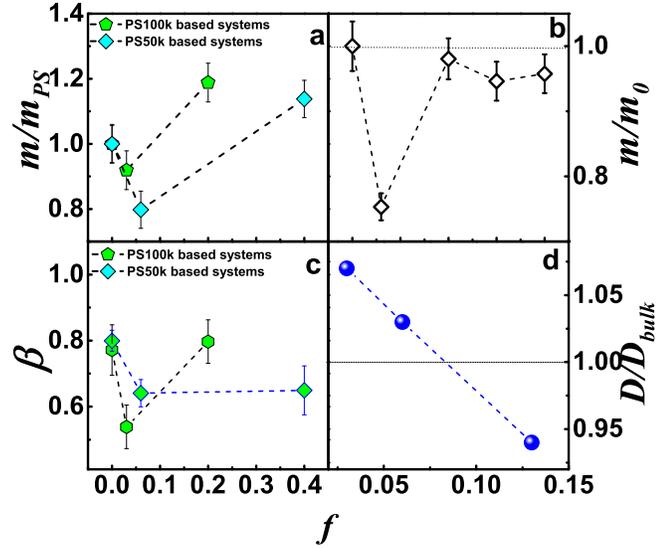

Figure 5: (a) Estimated fragility, $m$ normalized by $m_{PS}$, fragility of pure PS films as a function of entropic compatibility, $f$, from the temperature dependent viscosity is summarized, (b) Fragility, estimated from MD simulation, normalized with that of pure polymer system ($m_0$) as a function of $f$ showing a clear reduction in $m$ for smaller $f$, (c) Kohlrausch exponent as a function of $f$ showing a trend similar to that of $m$ and (d) Diffusivity of polymer chains near the nanoparticle surface, $D$, normalized to chain diffusivity far away from this interface, $D_{blk}$.

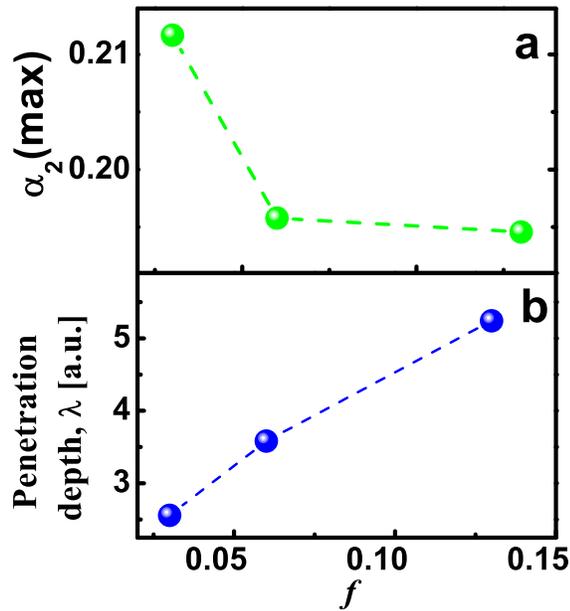

Figure 6: (a) The peak value of Non-Gaussian Parameter, $\alpha_2(max)$ as a function of $f$. (b) Penetration depth as a function of $f$

Thus our observation suggest that entropic PNC [27] with PGNP can show a reduction as well as enhanced fragility by just changing the entropic interaction parameter, $f$, without changing any direct nanoparticle-polymer or nanoparticle-nanoparticle interaction. In addition, Fig 5(c) also shows the variation of the average, $\beta$ for various values of $f$. The variation in $\beta$ with $f$ looks very similar to that of $m$. Usually, the smaller the value of $\beta$ larger is the heterogeneity in dynamics and in such cases, conventionally, the glass former has been identified as also being more fragile. However, in our case, we find that there is an intriguing anti-correlation.

Since the dynamics or viscosity in polymer nanocomposites is essentially determined by the PGNP-matrix IL, we explored the diffusivity of matrix chains in this region for various values of $f$ from MD simulations (Table S6). To do so we setup a system consisting of a single PGNP, with varied graft length for different $f$, fixed at the center of simulation box and surrounded by polymer chains. The interface and bulk region is demarcated using the matrix and graft monomer radial density profiles (Fig. S14). As we see from Fig 5(d), normalized (to bulk) IL diffusivity is enhanced at smaller $f$ and progressively reduces at larger $f$. This indicates that the interfacial diffusivity and fragility are inversely correlated. It is widely believed that for such entropic PNCs, smaller $f$ corresponds to a dewetting nanoparticle polymer interface [9,13,51] with attractive entropic interactions. As the interface becomes wetting (repulsive entropic interactions) by increasing $f$, the interface diffusivity reduces. Interestingly, the smaller value of $\beta$ corresponds to larger diffusivity (smaller IL width) and vice versa suggesting that, larger interfacial chain diffusivity could be related to a larger DH (i.e. smaller $\beta$). Therefore, these results suggests that the modified IL dynamics at the PGNP-PS interface is crucial in determining various dynamical parameters in these PNCs. To obtain further insight into the DH of the system and its relation with interfacial diffusivity as well as fragility, we have calculated non-gaussian parameter (NGP), $\alpha_2$, from MD simulation. The peak value of the NGP, $\alpha_2(max)$ (Fig. S18), for several values of $f$, are summarized in Fig. 6(a). It is quite evident that, $\alpha_2(max)$ decreases with increasing $f$. Existing reports have established a direct correlation between $\alpha_2(max)$ and DH of the liquid [26,52]. Therefore, the reduction in $\alpha_2(max)$ indicates a decrease in DH in the system with increasing $f$.

Since $f$ is not a microscopic parameter, we calculated the penetration depth, $\lambda$ of the matrix chains into grafted brush as function of $f$ which is shown in Fig. 6(b) [53]. It clearly shows a proportionality between $\lambda$ and $f$. We use this proportionality of $\lambda$ (eqn. 4 of SI) and $f$ to depict in Fig. 7 (a,b) the microscopic correlation of $m$, $\beta$ and $\alpha_2(max)$ with $\lambda$ and $D/D_{blk}$. While $\lambda$ and $D/D_{blk}$ has been obtained from MD simulation, and is difficult to extract in experiments, systematic trends are obtained for $m$ and DH in both simulations and experiments. While fragility, increases in with $\lambda$ and $D/D_{blk}$ DH represented by $1-\beta$, in experiments and $\alpha_2(max)$ in simulations decreases correspondingly. Hence, it is reasonable to conclude that IL microscopic parameters (quantified by $\lambda$ and $D/D_{blk}$) plays a considerable role in tuning the $m$ and DH of the PNC system. Importantly, our report provides an evidence of anti-correlation between fragility and dynamical heterogeneity which is contrary to most recent reports in other types or glasses or PNCs [5,25]. These results also points to the predominance of kinetic fragility in our entropic PNC systems which was alluded to occur in some amorphous polymers earlier [26] but has not been observed in most other PNC systems.

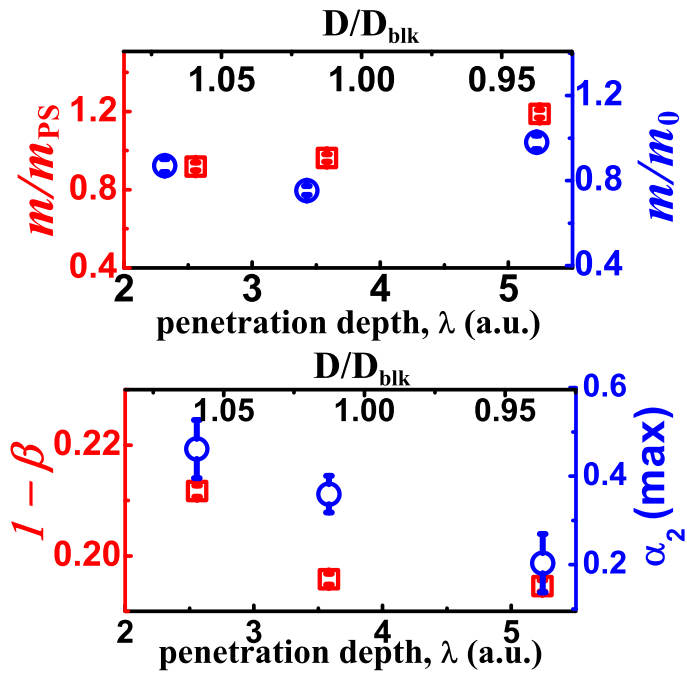

Figure 7: (a) Master curve showing fragility estimated from experiment ($m/m_{PS}$) as well as simulation ($m/m_0$), and (b) (1-$\beta$) and $\alpha_2(max)$ as a function of $\lambda$ and normalized IL diffusivity (D/D$_{blk}$).

# 4  Conclusion

In conclusion, we have studied the influence of nanoparticle-polymer interfacial layer dynamics on the fragility and DH of athermal PNC melts. The fragility of PNCs was found to increase with increasing entropic compatibility both in experiments and simulations. The DH on the other hand seemed to decrease with increasing $f$ suggesting an anti-correlation between fragility and DH, contrary to most earlier observations on glasses and polymers. A larger diffusivity of polymer chains was observed at the IL for smaller $f$, from simulations, which reduces with increasing $f$ and seems to drive the observed dynamical behavior. Finally, the extent of entropic interactions, as determined by $\lambda$, tunes the variation of both $m$ and DH. Our work illustrates the subtle correlations between interfacial layer dynamics and dynamical parameters of PNCs, which are critical in determining their flow and mechanical properties.

# Acknowledgements

Authors would like to thank the Department of Science and Technology (DST), India for the financial support for facilitating experiments at the P10 beamline in PETRA-III, DESY, Hamburg, Germany.Authors would like to thank the Department of Science and Technology (DST), India for the financial support for facilitating experiments at the P10 beamline in PETRA-III, DESY, Hamburg, Germany.

# SUPPLEMENTARY INFORMATION

In the main manuscript, we presented the role of interfacial entropic interaction (characterized by f) in tuning the fragility and the dynamical heterogeneity of polymer nanocomposites. Here, we provide the experimental and simulation details supporting the observations and interpretations discussed in the main manuscript.

# 1 Sample preparation

## 1.1 Synthesis of PGNPs

Thiol terminated polystyrene grafted gold nanoparticles (PGNP)were synthesized using the in-situ grafting-to method as was described by Lennox and coworkers [1]. Thiol terminated polystyrene (PST)and chloroauric acid (HAuCl4.3H2O)were dissolved separately in freshly distilled tetrahydrofuran (THF). The two solutions were then mixed and stirred for ca. 20 minutes to ensure homogeneous mixing of the components. Lithiumtriethylborohydride was added to reduce the chloroauric acid for facilitating the formation of gold nanoparticles. Simultaneously, the PST chains graft to the surface of the growing nanoparticle. Presence of grafted chains on the surface will inhibit further growth of nanoparticles before eventually seizing the growth.These nanoparticles as grown were precipitated by the addition of 3:1 (by volume) mixture of THF and ethanol, which is a good solvent for the free PST chains and a bad solvent for PGNPs. To ensure a complete removal of the ungrafted PST chains, PGNPs were dissolved in the 3:1 mixture of THF and ethanol and centrifuged. The procedure was repeated 4 times [2,3]. Finally, the precipitated PGNPs were dried in a desiccator.

## 1.2 Characterization of PGNPs

To determine the overall size of the PGNPs, we have performed small angle X-ray scattering measurements on dry powders of PGNPs in transmission geometry. Figure S1 shows the intensity profiles (intensity vs. transmission wave vector)of the two different PGNPs used in this study. A characteristic peak (indicated by an arrow mark in the respective panels) corresponding to the diameter of the PGNPs can be observed. The size of the 3k PST grafted PGNP was determined to be ca. 6nm and that for 20k PS grafted PGNP is around ca. 8.5 nm

## 1.3 Preparation of PNCs

PGNP and PS solutions were made separately and stirred for overnight, before mixing them in appropriate ratios to obtain a 0.5% by volume of gold core in the final solution [2,3]. The estimated total PGNP volume fraction at this core fraction (considering the total size obtained from SAXS) appears to be 2% for 3k PGNP and 6% for 20k PGNP. At this volume fraction, the grafted chain mass fraction in the solution for 3k PGNP becomes 13% and that for 20k PGNP becomes 35%. The mass fractions of matrix chains at this condition are 77% and 56% for 3k and 20k PGNP systems, respectively.

The polymer nanocomposite solution (PNC) was stirred overnight to ensure a homogeneous distribution of PGNPs. Solutions as obtained were spin coated on a freshly piranha treated silicon substrate. Four different PNC films were prepared along with bare PS films. Detail of the PNC samples is given in Table S1.

All the films were annealed at a temperature 145oC at a vacuum 10-6 mbar for 12 hours to erase possible preparation induced effects.

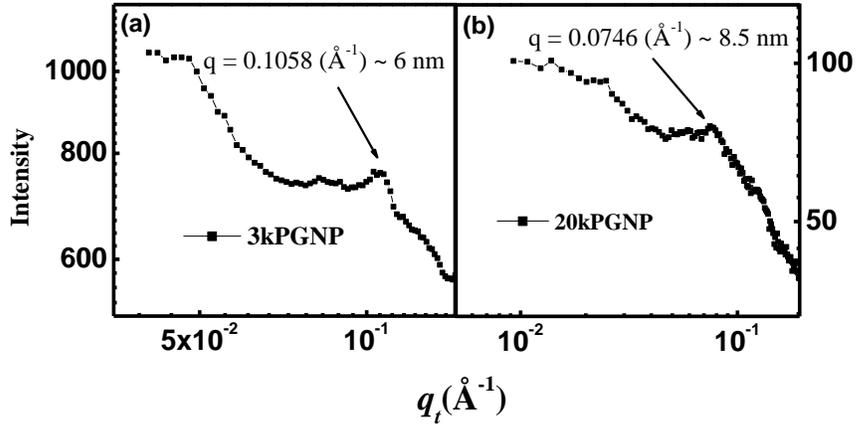

Figure S 1: Intensity profile along the transmission wave vector (qt) for PGNP with PS grafted molecular weight (a) 3k and (b) 20k is shown. From respective structure factor peaks (indicated by the arrow marks), we have determined the overall diameter of the PGNPs.

Table S 1: Important characteristics of the samples used in our studies

| Sample | Matrix chains, $W_m$ kDa | Grafted chains, $W_g$ kDa | $f$ | Grafting density, $\sigma$ chains/nm$^2$ |
|---|---|---|---|---|
| 3k50k | 50 | 3 | 0.06 | 1.7 |
| 20k50k | 50 | 20 | 0.4 | 1.3 |
| 3k100k | 100 | 3 | 0.03 | 1.7 |
| 20k100k | 100 | 20 | 0.2 | 1.3 |

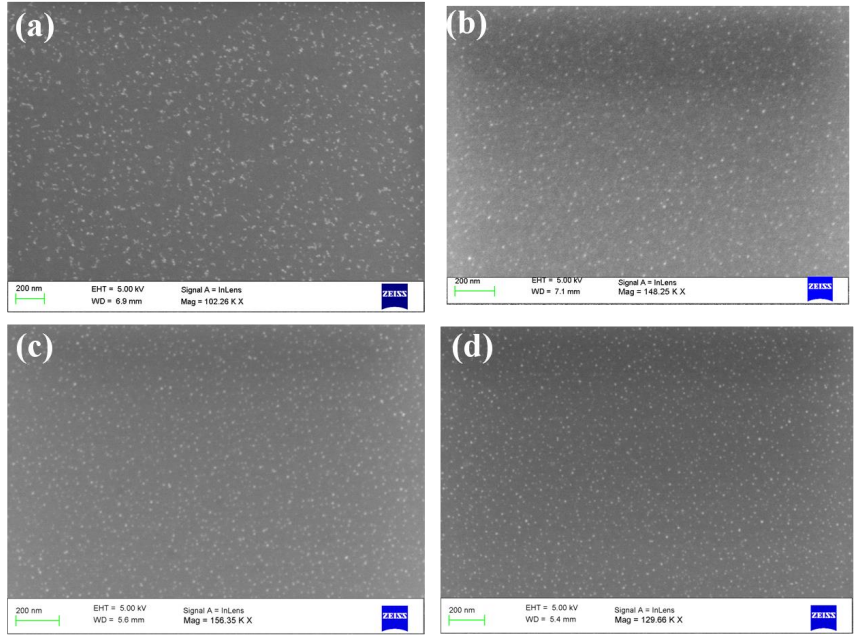

Figure S 2: Representative SEM micrographs of films with (a) $f = 0.03$, (b) $f = 0.06$, (c) $f = 0.2$ and (d) $f = 0.4$. A clear improvement in nanoparticle dispersion could be visualized with increase in $f$.

### 1.4 Entropic PNC

In this study, we have used polystyrene as graft and matrix molecular weight, hence, there is no enthalpy for mixing. Consequently, the only term that contributes to changes in free energy of the system is entropy, which depends on graft and matrix molecular weights and the grafting density [4]. Therefore, entropy decides the nature of dispersion of the particles and the resultant properties and hence, we called these PNCs as entropic PNC.

## 2 Dispersion state of PGNPs in annealed films

All the films were imaged using field emission scanning electron microscopy (SEM) to observe the nature of the dispersion state of PGNPs with different f. The SEM images of the films are presented in Fig. S2.

## 3 X-ray reflectivity profiles of the films for determining thickness

Thicknesses of the films were obtained from X-ray reflectivity (XR) profiles. The XR profiles of all the films are presented in Fig. S3. Thicknesses of the films were determined from the width of Kiessig-fringes ($\Delta q_z$) [2]. The obtained film thicknesses are summarized in the table S2.

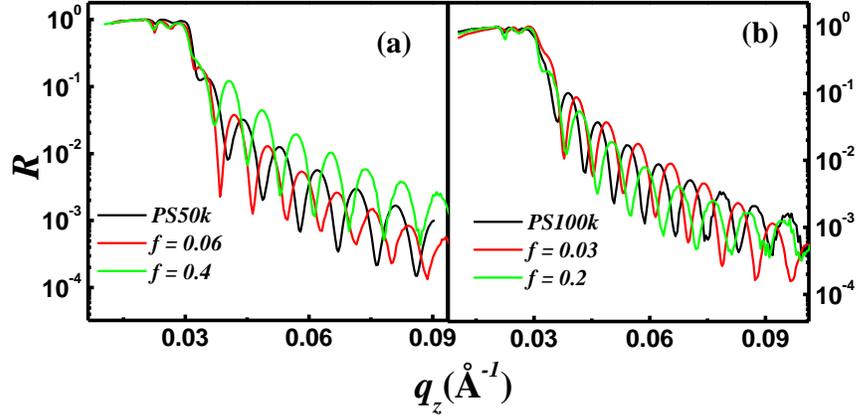

Figure S 3: XR profiles of PS50 based systems (left panel) and PS100k based systems (right panel) at temperature 403 K.

Table S 2: Thickness of the films obtained from XRR at temperature 403K

| Sample | Thickness (nm) |
|:---:|:---:|
| PS50k | 63.8 |
| 3k50k | 69.5 |
| 20k50k | 70.4 |
| PS100k | 73.8 |
| 3k100k | 70.4 |
| 20k100k | 67.6 |

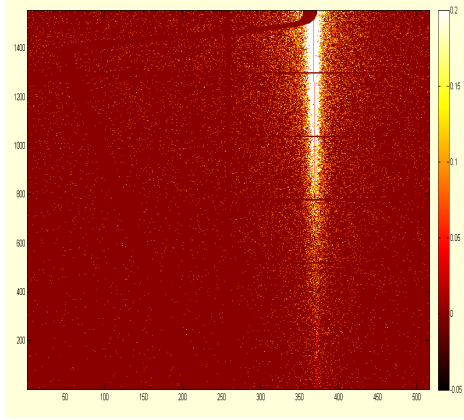

Figure S 4: A typical CCD image collected on PS100k sample in the reflection geometry at an incident angle, 0.15°, lesser than the film critical angle 0.16°.

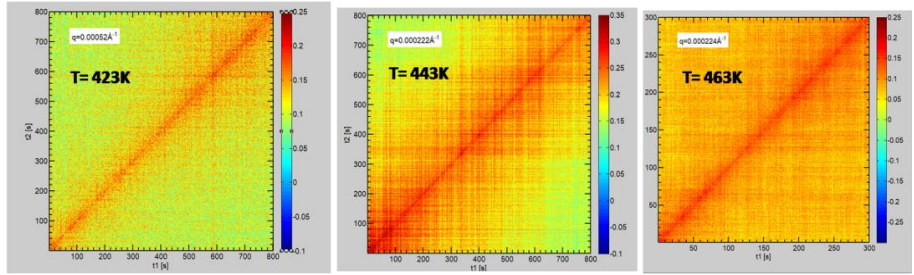

Figure S 5: Typical two time correlation function for samples (from left) 3k50k, 3k100k and 20k100k.

## 4 X-ray photon correlation spectroscopy and analysis

A typical CCD image of such scattering is shown in Fig. S4 [5]. In order to verify the equilibration of the system after annealing, we have studied the two time correlation functions during measurements which is given by

$$g(q, t_0, t_1) = \frac{\langle I(q,t_0)I(q,t_1)\rangle_\psi}{\langle I(q,t_0)\rangle_\psi \langle I(q,t_1)\rangle_\psi} \qquad (6)$$

where, $\langle ... \rangle_\psi$ denotes the average over detector pixels. Few typical plots of two time correlation functions are shown in Fig S5. The plots of the current systems show no variation in intensity with time (along the diagonal of the plot) indicating no time evolution of the system during measurements [6,7]. The samples therefore are in equilibrium. The $F(q_x, t)$ obtained from XPCS measurements are shown in Fig. S6 for all the samples at different temperatures for a particular $q_x$.

Extracted relaxation rate $\Gamma(q_x)$ for the PS100k based samples are shown in Fig. S7, as a function of $q_x h$.

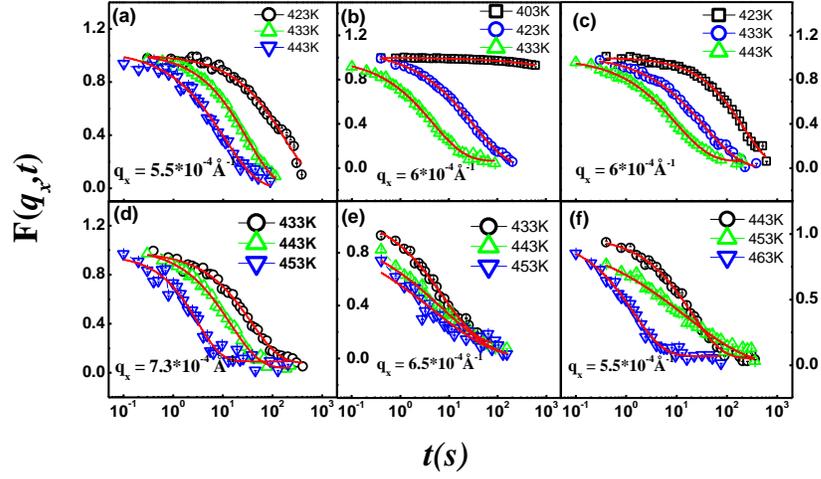

Figure S 6: ISF for (a) bare PS50k, (b) $f = 0.06$, (c) $f = 0.4$, (d) PS100k, (e) $f = 0.3$, and (f) $f = 0.2$ as a function of $t$ at different temperatures along with the fits (red solid lines).

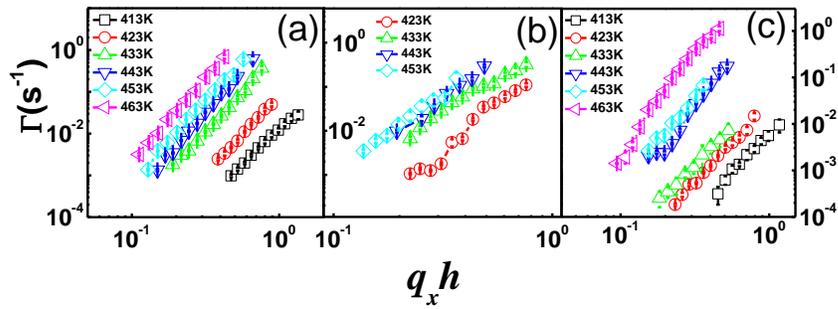

Figure S 7: Relaxation rate ($\Gamma$) as a function of $q_x h$ at different temperatures for PS100k (a), for $f = 0.03$ (b) and that for $f = 0.2$.

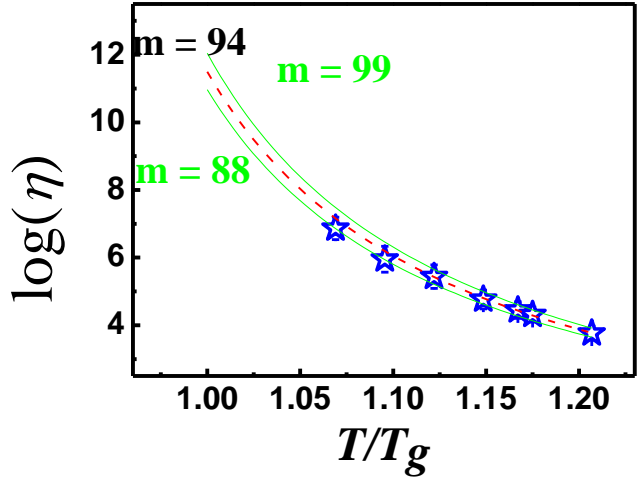

Figure S 8: Figure shows VFT fit (red dashed line) to the viscosity of PS50K sample and the extrapolation of the VFT fit from which the slope near $T_g$ ($T/T_g = 1$) is calculated to determine the fragility, $m$. The green lines show the fits at two extreme values of $T_g$ ($T_g$-error bar and $T_g$+error bar) which provides the actual error in estimating $m$.

## 5  VFT modeling to the viscosity of the films

The estimated viscosity (presented in main manuscript) indicates temperature dependence described by Vogel-Fulcher-Tammann (VFT) equation given by

$$\eta = \eta_0 e^{\frac{BT_0}{T-T_0}} \tag{7}$$

Using this equation, the viscosity was modeledand the fit was extrapolated till $T = T_g$. The Vogel-Fulcher temperature, $T_0$ was estimated from the glass transition temperature using the relation, $T_0 = T_g$ - 50. These estimated values of $T_0$ were used to fit the experimental data with eqn 2. The details of $T_g$ measurements are given in the next section. VFT fit for PS50k, extrapolated till $T = T_g$, is shown in Fig. S8. From such VFT fits, we have extracted the fragility, $m$, given by

$$m = \left.\frac{\partial log\eta}{\partial \left(\frac{T}{T_g}\right)}\right|_{T_g} \tag{8}$$

The slope of the $\eta$ vs $T$ plot near $T_g$ was used to determine $m$, which are summarized in Fig. 5 (a) of the main manuscript. In order to estimate errors in m, we have varied the value of parameter $T_0$ according to the error bars in experimentally estimated $T_g$. Corresponding VFT fits provide the extreme values of the fragility which gives the estimate of errors in $m$. A typical plot showing the three fits for sample PS50k where $T_g$ is varied within its error bar is shown in Fig. S8. This provides a larger error in $m$. We also include the errors obtained in the fit parameters, e.g. B in equation 2. This is how the errors are calculated and plotted in Fig. 5 (a) of the main manuscript.

## 6  $T_g$ measurement using force distance spectroscopy

Conventional Tg measurements for bulk PNCs, such as differential scanning calorimetry (DSC) can notbe performed on these PNC films. Hence, $T_g$ was measured using atomic force microscope based force-distance spectroscopy measurements performed at different temperatures [2,8]. We used a SiO2 cantilever with a curvature of tip radius $\sim$ 10 nm and resonance frequency $\sim$ 150kHz. Force experienced by the tip due to tip-sample interaction was recorded from the cantilever deflection as a function of tip-sample distance. The change in slope of the retrace curve was observed with increasing temperature (Fig. S9) as observed earlier [2]. The slope gives a measure of the combined stiffness of sample and cantilever. This slope has been plotted as a function of temperature in Fig. S10. The transition temperature can be estimated by modeling the data with a sigmoidal function and extracting the inflection point. All the calculated $T_g$ values are summarized in a table(see Table S3).

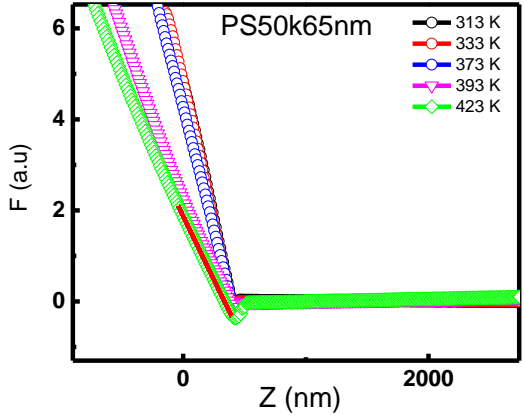

Figure S 9: Temperature dependent F-D retrace curve for PS film of thickness 65nm. Red curve on the T=423K data shows the slope on the retrace curve.

Table S 3: Estimated $T_g$ values using AFM force-distance spectroscopy

| Sample | $f$ | $T_g$ (°) |
|---|---|---|
| PS50k | 0 | 377 ± 2.5 |
| 3k50k | 0.06 | 376 ± 2.5 |
| 20k50k | 0.4 | 379 ± 5 |
| PS100k | 0 | 378 ± 2 |
| 3k100k | 0.03 | 362 ± 2.5 |
| 20k100k | 0.2 | 375 ± 4 |

Using AFM based force distance spectroscopy for measuring $T_g$ has been established earlier by ourselves and others [2, 8, 9, 10]. It has been shown that this approach gives reliable Tg values, which can be compared with other conventional techniques like DSC (in bulk) and ellipsometry (in thin films). For example, as shown in Fig. S8, within the experimental error, $T_g$ values of polystyrene films composed of $M_w = 100$ kg/mol and $M_w = 50$ kg/mol, yield values that are in good agreement with the bulk values obtained by DSC [11]. We have shown a comparison of Tg estimated from DSC on bulk PNC and AFM on films in table S4. Our result suggests that they are as accurate as DSC, reaffirming that AFM based experiments are one of the most reliable techniques for measuring $T_g$ on such films. Moreover, the accuracy obtained in $T_g$ determination is sufficient for fragility estimation which is the primary purpose of $T_g$ measurements in this manuscript.

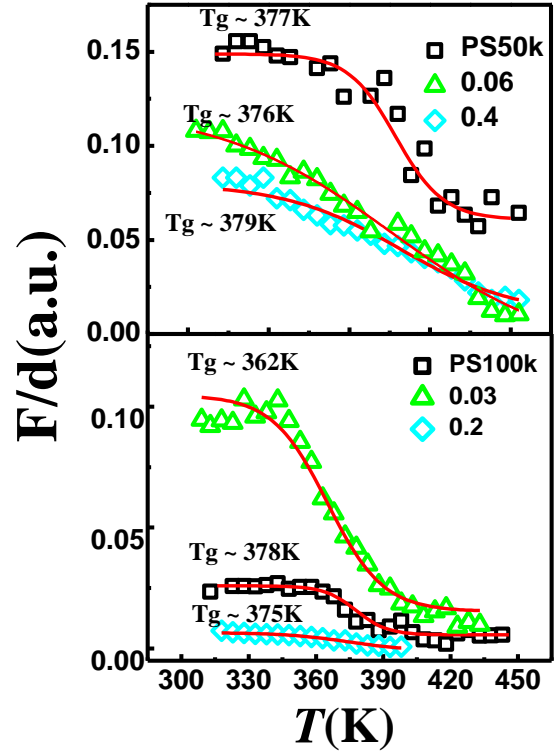

Figure S 10: Slope of retrace curves as a function of temperature for all PS100k based samples (top panel) and PS50k based samples (bottom panel) showing the variation of $T_g$ with $f$ compared to bare PS films

Table S 4: Comparison of $T_g$ values estimated from DSC and AFM

| Sample | AFM $T_g$(K) | DSC $T_g$ |
|--------|--------------|-----------|
| PS50k  | 377          | 376       |
| 20k50k | 379          | 375       |
| PS100k | 378          | 378       |
| 20k100k| 375          | 376       |

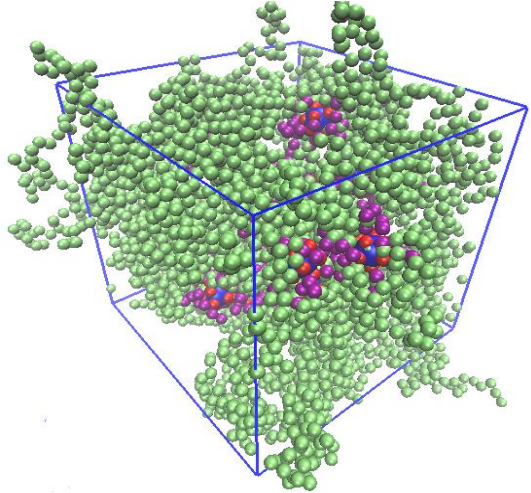

Figure S 11: Visual Molecular Dynamics (VMD) [19] Snap shot of a typical composite system. Nanoparticle (Blue), Tethered (Red), Grafted(Purple), Matrix monomers (Green). The graft chain here hasN=3monomers. The density of matrix is reduced than actual value to show different components.

# 7 Calculation of viscosity and fragility from MD simulation

The model nanocomposites were prepared by generating a mixture of amorphous linear polymer chains and polymer-grafted nanoparticles (PGNP) in a cubical simulation box with periodic boundary conditions on all three sides. The matrix and grafted polymers were modeled as coarse-grained bead-spring chains, using the finite extensible nonlinear elastic potential with standard values [12]. All monomers were chemically identical with reduced mass M = 1.0 and diameter $\sigma = 1$. The degree of polymerization for the matrix chains was fixed at $M_m = 50$, while the graft length Mg was varied from 3 to 20 such that the ratio $f = M_g/M_m$ varied from 0.06 to 0.4. The nanoparticles were represented as uniform spheres of size $D = 4\sigma$ and mass proportional to D3. Polymer-grafted nanoparticles were then constructed by uniformly grafting polymer chains onto the surface of each nanoparticle such that the grafting density $\sigma_g = 0.7/\sigma^2$. A typical snapshot of one such system is shown in Fig. S11. The number of polymer chains ($N_m$) and nanoparticles ($N_n$) in the system were chosen such that the volume fraction of nanoparticles $\phi_n = D^3 N_n/(d^3 N_m + D^3 N_n)$= 0.05. The initial configurations were generated by randomly placing all particles within the box and any overlaps between monomers are removed by applying a soft cosine potential that slowly pushes the particles apart until the distance between their centers is equal to sum of the two radii [12]. The core and tethered beads of PGNP were treated as a single rigid body to prevent sliding of grafted chains on the particle surface. The pair-wise interactions between all monomers in the system were defined using a shifted Lennard-Jones (LJ) potential of the form [12]

$$E = 4\epsilon[(\frac{\sigma}{r})^{12} - (\frac{\sigma}{r})^6] - E_{r_c} \qquad (9)$$

truncated at $r_c = 2.5\sigma$, where $\sigma$ is the monomer size, $\epsilon$ is the energy scale, and $E_{(r_c)}$ is the potential at the cutoff distance $r_c$. The polymer-nanoparticle interaction was the LJ potential with a cutoff distance $r_c = D/2 + \sigma$, while two nanoparticles interacted with a purely repulsive LJ interaction. All thermodynamic quantities are expressed in reduced units that are convenient in molecular simulations. The systems were equilibrated by running isobarically at P*= 0 that represents atmospheric pressure

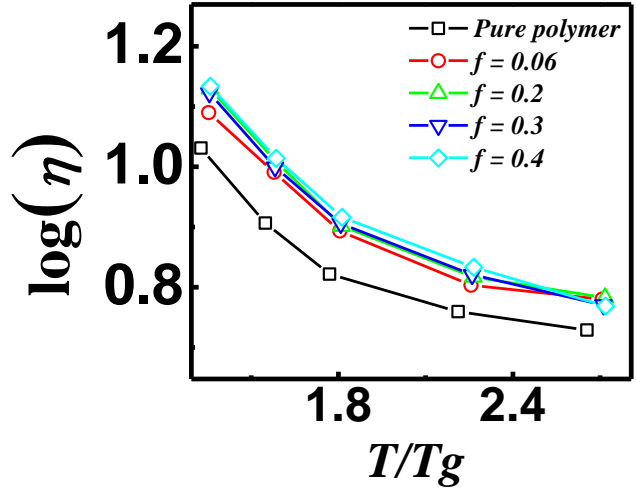

Figure S 12: Figure shows viscosity calculated as a function of temperature for neat polymer (red solid symbol) and different $f$ values as labeled in the figure.

[14], and then at constant volume until the chains move their own size using the LAMMPS simulation package [15]. The desired temperature was maintained using a Nose-Hoover thermostat with a damping parameter of 0.1. The specific volume of the system as a function of temperature to estimate $T_g$ was collected from subsequent production runs. The viscosity of the composite was estimated following two procedures at different temperatures. At high temperatures ($T/T_g > 1.55$), the zero shear rate viscosity was calculated from the stress autocorrelation function using the Green-Kubo relation [16],

$$\eta = \frac{V}{k_B T} \int_0^\infty \langle \sigma(t) \sigma(0) \rangle \, dt \qquad (10)$$

where $V$ is the volume, $T$ is the temperature, and $k_B$ is the Boltzmann's constant. At lower temperatures ($T/T_g < 1.55$), the SLLOD equations of motion, which adopts the transpose of the qp-DOLLS tensor [17] were integrated at a strain rate using a timestep of $\delta t = 0.005\tau$, where $\tau = \sqrt{(M\sigma^2/\epsilon)}$, where $M$ is the mass of a chain monomer, such that the diagonal components of the pressure tensor equal to zero [18]. The shear viscosity is calculated using $\eta = <P_{xz}>/\dot{\gamma}$, where $<P_{xz}>$ is the xz component of the pressure tensor along the flow and gradient directions, respectively. The viscosity values obtained from the two temperature ranges using two different methods have been merged to get a larger range of temperature. All results reported here are obtained by averaging values from three independent simulations. Calculated viscosity with varying temperature is shown in Fig. S12. The temperature dependent viscosity was fitted with VFT equation keeping the parameter $log(\eta_0)$ fix at a particular value (0.534 obtained from pure polymer data fitting) to reduce the number of fit parameters. Fragility of these systems were calculated using this method and summarized in Fig 4(a) of main manuscript.

Table S 5: System details

| System | Grafted chain length | Matrix chain length | $f$ |
|---|---|---|---|
| 1 | 10 | 300 | 0.03 |
| 2 | 18 | 300 | 0.06 |
| 3 | 40 | 300 | 0.13 |

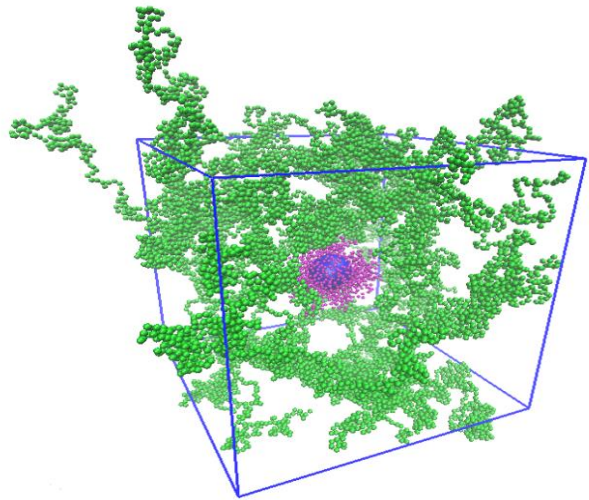

Figure S 13: VMD [19] snapshot for $f = 0.06$ system. The Nanoparticle (blue), Grafted chains (Pink) matrix chain monomers (Green). The graft chain here has $N = 18$ monomers. In this figure, the density of matrix as well as graft is reduced than actual value to show different components clearly

# 8 Calculation of diffusion coefficient

## 8.1 System details

The polymer and polymer grafted nanoparticles were modeled as described in previous section. However, in this case we have just a single PGNP fixed at the center of simulation box.In order to study systems with different $f(= N_grafted/N_matrix)$, we varied the number of monomer units of the grafted chains while keeping the matrix chain length fixed at $N = 300$. So, for higher $f$ values we have a larger PGNP. Therefore, in order to have a well-defined bulk and interfacial region, the number of matrix chains was made higher for larger $f$ systems. We have 600, 350 and 200 matrix polymer chains for $f = 0.13, 0.06$ and $0.03$ respectively which are summarized in table S5. Figure S13 shows one such system with lesser density than actual value.

## 8.2 Equilibration

The systems were equilibrated at constant volume and temperature $1\epsilon/$k. The thermostat and damping parameter is same as described in previous section. The equilibration is achieved by running each simulation for $1.5*10^6$ or 300 million MD time steps each time steps being $0.005\tau$ . The equilibration is checked by calculating the mean square internal displacement (MSID) [20]. The saturation of MSID

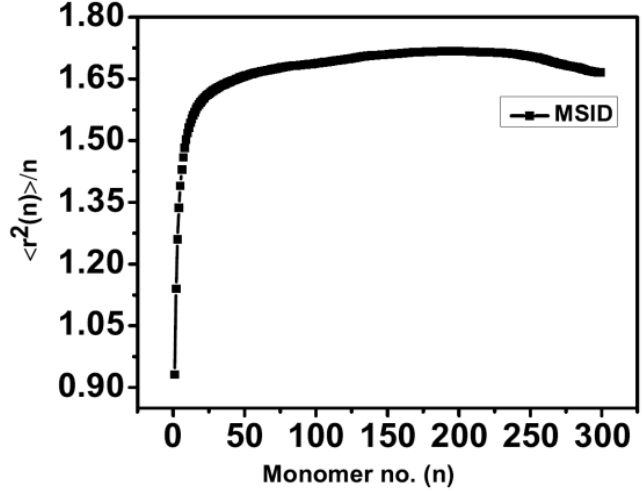

Figure S 14: MSID for matrix chains. Saturation of MSID was taken to be the signature of equilibration [15].

is taken as signature of equilibration. MSID is shown in Fig. S14.

### 8.3 Radial density of grafted monomers

A chain is assigned to one of the regions according to the position of its center of mass at time $t_1$ in Eqn. 6 [21]. The interfacial region is defined as the region lying within a radial distance of $r \sim R_n + h_g + R_g/2$. Where, $R_n$, $h_g$ and $R_g$ are the radius of PGNP core, brush height of grafted chains and the radius of gyration of matrix polymers, respectively. The brush height $h_g$ is defined as [22],

$$\langle h \rangle^{1/2} = \left( \frac{\int_{R_n}^{\infty} 4\pi r^2 (r - R_n)^2 \rho(r) dr}{\int_{R_n}^{\infty} 4\pi r^2 \rho(r) dr} \right)^{1/2} \quad (11)$$

where, $\rho(r)$ is the radial density of grafted monomers. The bulk and interfacial regions are shown in the density profiles (Fig. S15) of matrix and graft monomers.

### 8.4 Penetration depth, $\lambda$

Penetration of matrix chains into the graft polymer brushes were quantified using the parameter, penetration depth, $\lambda$, given by

$$\lambda = \left( \frac{\int_{0}^{h_g} 4\pi r^2 (r - R_n)^2 \rho_m(r) dr}{\int_{0}^{h_g} 4\pi r^2 \rho_m(r) dr} \right)^{1/2} \quad (12)$$

where $\rho_m(r)$ is the is the matrix radial density profile from the center of nanoparticle, $r$ is the radial distance, $R_n$ is the radius of nanoparticle and $h_g$ is the graft chain brush height.

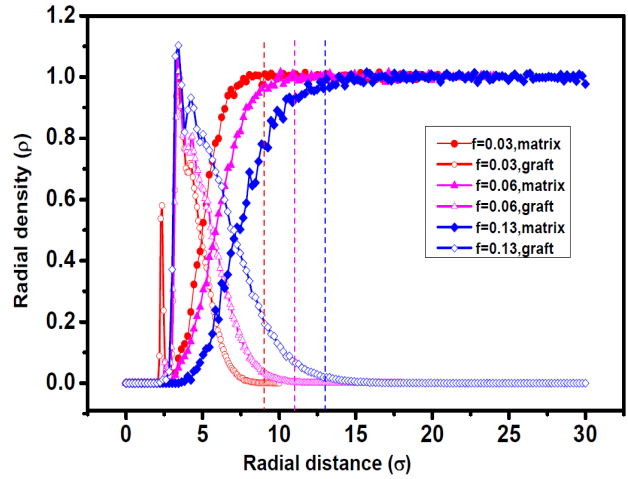

Figure S 15: Radial density profiles for graft monomers (open symbols) and Matrix monomers (solid symbols) showing bulk and interfacial region. The right side region of the vertical dashed line corresponds to bulk and the left side corresponds to interfacial region. The dashed lines are coloured such that a dashed line of particular colour represents interface boundary for a density profile with that colour.

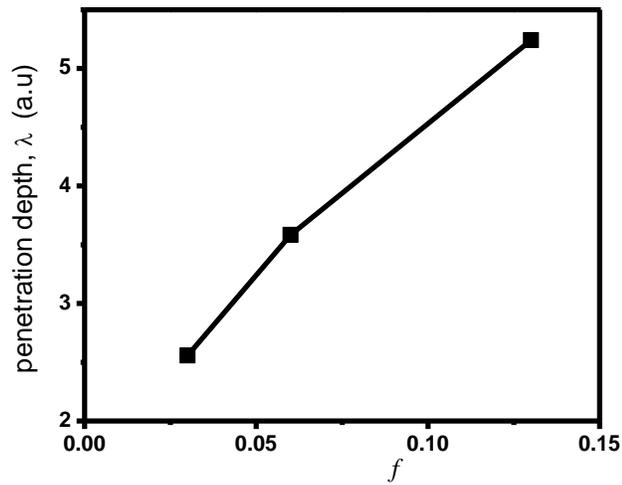

Figure S 16: Penetration depth calculated for different $f$s

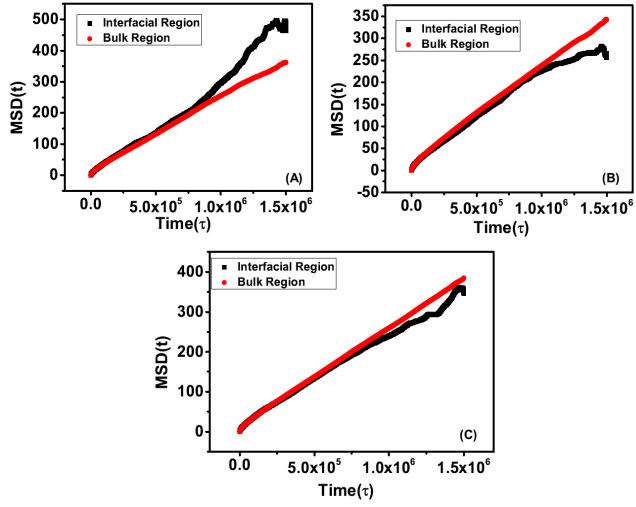

Figure S 17: MSD for different systems. (A) $f = 0.03$ (B) $f = 0.06$ and (C) $f = 0.13$.

Table S 6: Diffusivity values

| f | Interface diffusivity (x$\frac{10^{-4}}{6}\sigma^2/\tau$) | Bulk diffusivity (x$\frac{10^{-4}}{6}\sigma^2/\tau$) |
|---|---|---|
| 0.03 | 2.65 | 2.47 |
| 0.06 | 2.24 | 2.17 |
| 0.13 | 2.33 | 2.47 |

## 8.5 Diffusivity calculation

The equilibrated systems are run for further 300 million time steps ($\delta t = 0.005$) in order to calculate the diffusivity. The MSD of center of mass is given by [13],

$$g_3(t) = \langle [r_{cm}(t_2) - rcm(t1)]^2 \rangle \tag{13}$$

where $r_{cm}(t)$ is the center of mass position at time $t$ and $\langle . \rangle$ denotes time and ensemble averaging. The diffusivity is calculated using Einstein's equation,

$$g_3(t) = 6Dt \tag{14}$$

For a given system, we calculated the MSD (Fig. S17) for two different regions, interfacial and bulk region, in the simulation box. The values of diffusivity at interface and bulk regime are summarized in table S6.

## 8.6 Calculation of $T_g$ of the simulation systems

For the calculation of fragility, $T_g$ is estimated from the temperature dependence of specific volume. It is calculated as the point at which the slope of the specific volume vs temperature curve changes. One typical plot showing $T_g$ of the system is shown in Fig. S18.

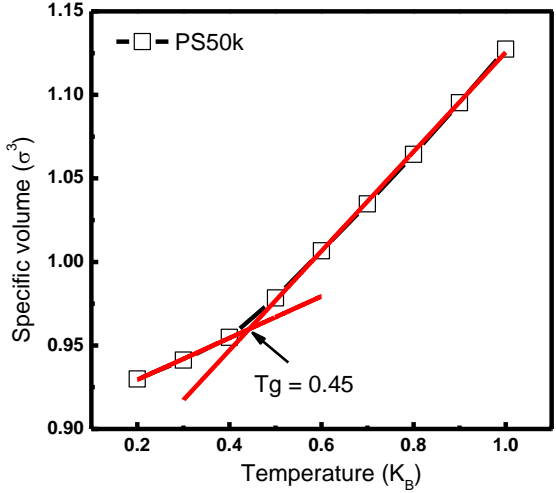

Figure S 18: Variation of specific volume for a bulk polystyrene ($M_w \sim 50$ kDa) system generated using coarse grained molecular dynamics simulation. The point at which slope change occurs is taken as $T_g$.

### 8.7 Non-Gaussian parameter

The Non-Gaussian parameter was calculated using [23],

$$\alpha_2(t) = \frac{3\langle \Delta r(t)^4 \rangle}{5\langle \Delta r(t)^2 \rangle^2} - 1 \qquad (15)$$

where $\langle .. \rangle$ denotes time and ensemble averaging over all the monomers in the system. $\Delta r(t)$ is the displacement of a monomer in a time interval $t$. The NGP for all the system is shown in Fig. S19. The peak value of $\alpha_2$, that characterizes the microscopic dynamics of the system, is summarized in Fig. 5(d) in the main manuscript.

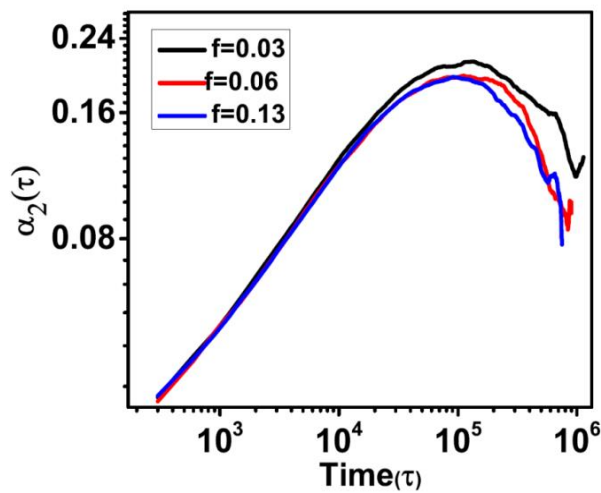

Figure S 19: Non-Gaussian parameter for different systems